# Top-r Influential Community Search in Bipartite Graphs

Yanxin Zhang*
yzhang2879@wisc.edu
University of Wisconsin - Madison

Zhengyu Hua*
hzysgg@njust.edu.cn
Nanjing University of Science and Technology

Long Yuan†
longyuan@whut.edu.cn
Wuhan University of Technology

Zi Chen
zichenscs@gmail.com
Wuhan University of Technology

## ABSTRACT

Community search on bipartite graphs, especially influential community detection, has received significant attention. Existing studies use minimum vertex weights, inadequately reflecting true community influence when some vertices have low weights. To address this, we introduce the $(\alpha, \beta)$-influential community model based on the average vertex weights from both layers, providing a more comprehensive influence measure. Given the NP-hardness of accurately identifying top-$r$ communities, we propose an exact recursive algorithm enhanced by a slim tree structure and upper-bound techniques to improve efficiency. Additionally, we develop a greedy approximate algorithm with $O((n+m)+m\log n)$ complexity, further optimized by a pruning strategy. Experiments on 10 real-world graphs demonstrate the effectiveness and efficiency of our proposed algorithms.

**VLDB Workshop Artifact Availability:**
The source code, data, and/or other artifacts have been made available at https://github.com/yanxinzhangcs/Top-r-Influential-Community-Search-in-Bipartite-Graphs.

## 1 INTRODUCTION

Many real-world relationships can be represented as bipartite graphs, including customer-product networks[28], user-page networks[1], gene co-expression networks[11], and collaboration networks[14]. With the growth of bipartite graph applications, extensive research has addressed fundamental problems related to their management and analysis, notably community search.

Community search in bipartite graphs traditionally emphasizes structural cohesiveness using models such as the $(\alpha, \beta)$-core[4, 7, 18, 19], bitruss[27, 29, 30], biplexyuan2025efficient, and biclique[5, 21, 33]. Applications include anomaly detection[22], personalized recommendation[13], and gene expression analysis[23]. Other studies integrate vertex attributes into community detection[6, 8, 10, 32].

*Both authors contributed equally to this research.
†Corresponding author.

**Motivations.** Existing community search typically overlooks vertex weights, prompting studies into influential community search that connect vertex importance to community influence[2, 9, 12, 15–17, 20, 24–26, 31, 34, 35]. For instance, [34] introduces a model where community influence is measured by the minimum vertex weights across layers, ensuring high overall vertex influence. However, a single low-weight vertex significantly reduces measured community influence. To overcome this, we propose an $(\alpha, \beta)$-*influential community* model based on the $(\alpha, \beta)$-core structure, defining community influence as the sum of average weights from both vertex layers. An $(\alpha, \beta)$-influential community is thus a maximal connected $(\alpha, \beta)$-core not included in another $(\alpha, \beta)$-core of equal influence.

**Applications.** Our model has diverse real-world applications:

- **Team Formation.** In developer-project graphs, developer weights represent ability, and project weights represent importance. Our model can help identify cohesive teams comprising skilled developers associated with significant projects.
- **Movie Recommendation.** User-movie networks, where user weights indicate activeness and movie weights indicate ratings, allow recommendations of quality movies liked by active users.
- **Fraud Detection.** Customer-item graphs from platforms like Amazon and Alibaba, with vertex weights indicating transaction and purchase frequencies, can identify potentially fraudulent communities involving suspicious customers or items.

**Challenges.** Previous minimum-weight-based influential models allow efficient, linear-time solutions through effective pruning. However, using average vertex weights as community influence renders the problem NP-hard, posing significant computational challenges.

**Our Approach.** We propose exact and approximate solutions to address these challenges. Initially, we present an exact recursive algorithm exploring all subgraphs. Enhancements include using a slim-tree structure to reduce search width and an upper-bound pruning strategy to reduce search depth. Given the NP-hardness, exact searches remain computationally intensive, motivating our development of a greedy approximate algorithm balancing accuracy and efficiency.

**Contributions.** This paper makes the following contributions:

- **New community model.** Introduces the $(\alpha, \beta)$-influential community model integrating vertex importance and structural cohesiveness.
- **Exact algorithms.** Proposes three exact algorithms: a basic recursive approach, a slim-tree optimization, and an upper-bound pruning approach.

- **Approximate algorithms.** Develops two efficient approximate algorithms, including a greedy strategy (complexity $O((n+m)+m\log n)$) and its pruned improvement.
- **Extensive experiments.** Conducts thorough experiments on 10 real-world datasets, validating model effectiveness and algorithm efficiency.

# 2 PROBLEM DEFINITION

An undirected vertex-weighted bipartite graph $G = (U, V, E)$ is a graph consisting of two disjoint sets of vertexs called layers $U$ and $V$ such that every edge from $E \subseteq U \times V$ connects one vertex of $U$ and one vertex of $V$. We use $U(G)$ to denote the set of vertices in the upper layer, $V(G)$ to denote the set of vertices in the lower layer, $E(G)$ denotes the set of edges. We denote the number of vertexs in $U$ and $V$ as $n_u$ and $n_v$, the total number of vertexs as $n$ and the number of edges in $E(G)$ as $m$. The set of neighbours of a vertex $u$ in $G$ denotes $N_G(u)$, and the degree of $u$ is denoted as $\deg(u, G) = |N_G(u)|$. Moreover, in each vertex $u \in U(G) \cup V(G)$ has a weight $w(u)$.

*Definition 2.1.* **($(\alpha, \beta)$-core)** Given a bipartite graph $G$ and two integers $\alpha$ and $\beta$, the $(\alpha, \beta)$-core of $G$, denoted by $C_{\alpha,\beta}$, consists of two vertex sets $U' \subseteq U(G)$ and $V' \subseteq V(G)$ such that the bipartite subgraph $G'$ induced by $U' \cup V'$ is the maximal subgraph of $G$ in which all the vertexs in $U'$ have degree at least $\alpha$ and all the vertexs in $V'$ have degree at least $\beta$, i.e., $\forall u \in U'$, $\deg(u, G') \geq \alpha$ , and $\forall v \in V', \deg(v, G') \geq \beta$.

*Definition 2.2.* **(Influence value of a community)** Given an induced subgraph $S$ of a bipartite graph $G$, its influence value $f(S) = f_U(S) + f_V(S)$, where $f_U(S)$ is the average value of the weights of all vertices in upper layer (i.e., $f_U(S) = \sum_{u \in U(S)} w(u)/|U(S)|$), $f_V(S)$ is the average value of the weights of all vertices in lower layer (i.e., $f_V(S) = \sum_{v \in V(S)} w(v)/|V(S)|$).

*Definition 2.3.* ($(\alpha, \beta)$-**influential community**) Given a bipartite graph $G = (U, V, E)$ and two integers $\alpha$ and $\beta$, an $(\alpha,\beta)$-influential community is an induced subgraph $S$ of $G$ that meets all the following constraints.

- **Connectivity**: $S$ is connected;
- **Cohesiveness**: Each vertex $u \in U(S)$ satisfies $\deg(u, S) \geq \alpha$ and each vertex $v \in V(S)$ satisfies $\deg(v, S) \geq \beta$;
- **Maximality**: there does not exist another induced subgraph $S'$ of $G$ such that (1) $S'$ satisfied connectivity and cohesiveness constraints, (2) $S'$ contains $S$, and (3) $f(S') = f(S)$.

**Problem statement.** Given a bipartite graph $G = (U, V, E)$ and three integers $\alpha$, $\beta$ and $r$, the problem is top-$r$ influential community search to compute $r$ $(\alpha,\beta)$-influential communities in $G$ with the highest influence value.

*Example 2.4.* Consider the bipartite graph $G$ in Figure 1. The weight of each vertex is shown as the circled value. There exist three (2, 2)-communities which are marked with three different colors ($H_1$, which contains $\{u_1, u_2, u_3, v_1, v_2, v_3\}$. $H_2$, which contains $\{u_2, u_3, v_2, v_3\}$. $H_3$, which contains $\{u_6, u_7, v_7, v_8\}$). However, $H_4 = \{u_1, u_3, v_1, v_2\}$ is not a (2,2)-influential community as $f(H_4) = f(H_1)$ =4 and $H_4 \subseteq H_1$ which does not satisfy the maximality constraint.

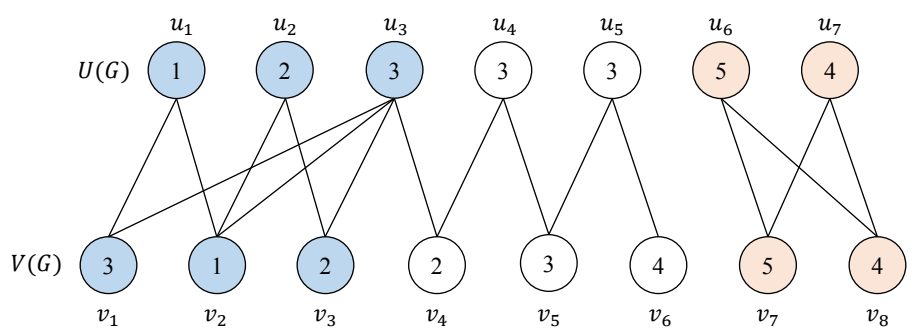

**Figure 1: A bipartite graph $G$**

**Problem Hardness.** The top-$r$ $(\alpha,\beta)$-inflential community search problem is NP-hard, which is shown as follows:

Theorem 2.5. *The top-r $(\alpha,\beta)$-influential community search problem is NP-hard.*

**Proof:** We prove this theorem based on the NP-Hardness of top-$r$ $k$-influential community search [26]. Given an unipartite vertex-weighted graph $G$, a $k$-influential community $C$ is a connected $k$-core, and there does not exist another subgraph $C'$ of $G$ such that: (1) $C'$ is a connected $k$-core, (2) $C$ is a subgraph of $C'$ and the influence value of $C$ is the same as that of $C'$, i.e., $f(C') = f(C)$, where $f(G) = \sum_{u \in V(G)} w(u)/|V(G)|$. Top-$r$ $k$-influential community search aims to compute $r$ $k$-influential communities with the highest influence value in $G$. For a given vertex-weighted unipartite graph $G$, we can transfer $G$ into a bipartite graph $G'$ as follows: for each vertex $v \in V(G)$, there exist two mirror vertices $u' \in U(G')$ and $v' \in V(G')$ with $w(u') = w(v') = w(v)$. For each edge $(u, v) \in E(G)$, there exists two edges $(u', v') \in E(G')$ and $(u'', v'') \in E(G')$, where $u'/v'$ and $u''/v''$ are the mirror vertices of $u$ and $v$ in $U(G')/V(G')$. It is clear that each $k$-influential community corresponds to a $(k, k)$-influential community in $G'$, which means the top-$r$ $k$-inflential community search problem in $G$ can be reduced to the problem of top-$r$ $(\alpha,\beta)$-influential community search problem in $G'$. As the top-$r$ $k$-inflential community search problem is NP-hard, our top-$r$ $(\alpha,\beta)$-influential community search problem is also NP-hard.

# 3 EXACT ALGORITHMS

In this section, we focus on developing exact algorithms for the problem. Our algorithms are based on recursion that derive the optimum result.

## 3.1 The Basic Algorithm

Lemma 3.1. *For any graph G, each maximal connected component of the maximal $(\alpha,\beta)$-core of G is an $(\alpha,\beta)$-influential community.*

**Proof:** The proof can be easily obtained by definition.

**Algorithm.** Algorithm 1 outlines the basic recursive procedure. Starting with inputs $G$, $\alpha$, $\beta$, and $r$, it initializes a priority queue $S$ ordered by community influence. The algorithm identifies the maximal $(\alpha, \beta)$-core and its connected components (lines 6-7). It recursively explores subgraphs potentially containing optimal solutions (lines 18-21) and checks feasibility and maximality conditions (lines 9-17). When a feasible solution $h$ with higher influence is found, it ensures maximality by comparing against existing communities in $S$ and updates accordingly (lines 10-17).

*Example 3.2.* Figure 2 illustrates Algorithm 1 finding the top-1 (2,2)-influential community. The algorithm identifies the maximal

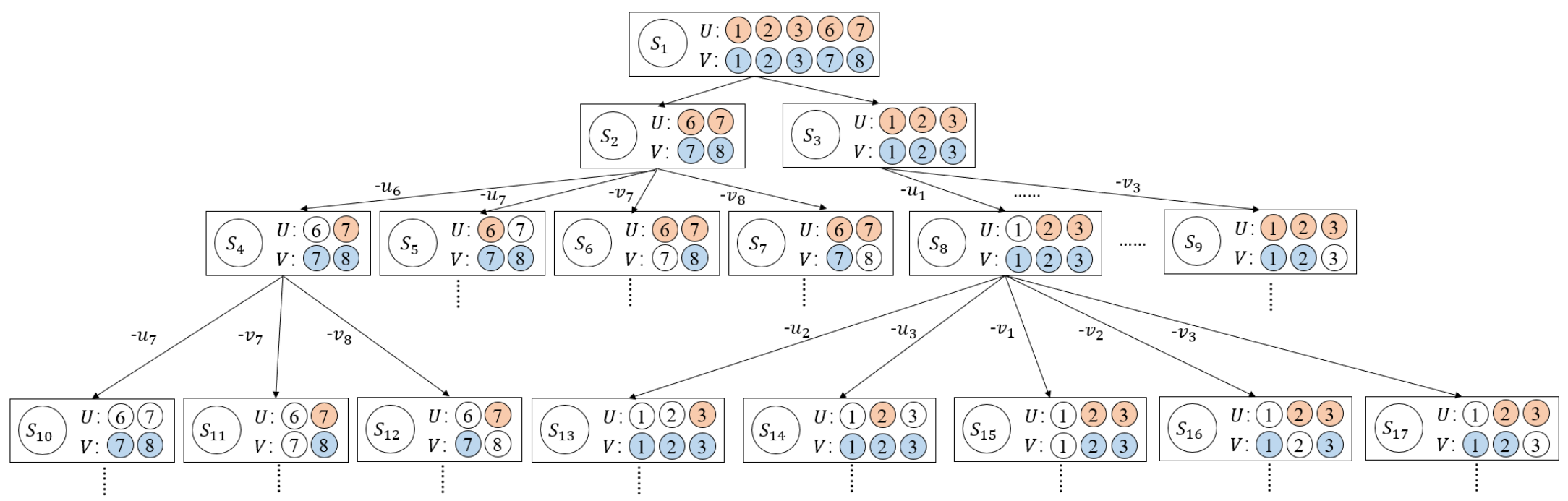

**Figure 2: An example of basic search**

**Algorithm 1:** Basic Algorithm

**Input:** $G = (U, V, E)$, $\alpha$, $\beta$, $r$

**Output:** Top-$r$ ($\alpha$,$\beta$)-influential communities

1 **Function** `Main()`:

2 $S \leftarrow \emptyset$; `Find`($G$); **return** $S$;

3 **Procedure** `Find`($G$):

4 $G \leftarrow$ maximal $(\alpha, \beta)$-core of $G$; $H \leftarrow$ connected components of $G$; $h_{min} \leftarrow r$-th largest community in $S$; **foreach** $h \in H$ **do**

5 **if** $f(h) > f(h_{min})$ **then**

6 $flag \leftarrow true$; **foreach** $h' \in S$ **do**

7 **if** $f(h) = f(h')$ ***and*** $h \subseteq h'$ **then**

8 $flag \leftarrow false$; **break;**

9 **if** $flag$ **then**

10 $S \leftarrow (S \setminus h_{min}) \cup h$;

11 **foreach** $u \in U$ **do**

12 `Find`($h \setminus \{u\}$);

13 **foreach** $v \in V$ **do**

14 `Find`($h \setminus \{v\}$);

$(\alpha, \beta)$-core ($S_1$), splits into connected components ($S_2$, $S_3$), and recursively searches these subgraphs. Only part of the search process is shown to highlight the benefits of subsequent algorithms.

Theorem 3.3. *Algorithm 1 correctly identifies the top-r $(\alpha, \beta)$-influential communities.*

**Proof:** Algorithm 1 recursively explores subgraphs, verifies cohesiveness via the maximal $(\alpha, \beta)$-core (line 6), and ensures connectivity by examining components (line 7). It maintains and updates the top communities, ensuring maximality (lines 12-15).

Theorem 3.4. *Algorithm 1 has a time complexity of $O((m + n + |H|r) \cdot 2^n)$.*

**Proof:** Each recursion level involves maximal core identification ($O(m)$), component detection ($O(m + n)$), and updating the community set ($O(|H|r)$). With binary inclusion/exclusion choices per vertex, total complexity is $O((m + n + |H|r) \cdot 2^n)$.

**Algorithm 2:** Slim Tree Structure

**Input:** $G = (U, V, E)$, $\alpha$, $\beta$, $r$

**Output:** Top-$r$ ($\alpha$,$\beta$)-influential communities

1 **Function** `Main()`:

2 $S \leftarrow \emptyset$; `Find`($G$); **return** $S$;

3 **Procedure** `Find`($G$):

4 $G \leftarrow$ maximal $(\alpha, \beta)$-core of $G$; $H \leftarrow$ connected components of $G$; $h_{min} \leftarrow r$-th largest community in $S$; **foreach** $h \in H$ **do**

5 **if** $f(h) > f(h_{min})$ **then**

6 $flag \leftarrow true$; **foreach** $h' \in S$ **do**

7 **if** $f(h) = f(h')$ ***and*** $h \subseteq h'$ **then**

8 $flag \leftarrow false$; **break;**

9 **if** $flag$ **then**

10 $S \leftarrow (S \setminus h_{min}) \cup h$;

11 **while** $U \neq \emptyset$ **do**

12 $u \leftarrow U.front$; $U \leftarrow U \setminus \{u\}$; $G' \leftarrow h \setminus \{u\}$; $G_{\alpha,\beta} \leftarrow$ maximal $(\alpha, \beta)$-core of $G'$; $U \leftarrow U \setminus (G' \setminus G_{\alpha,\beta})$; `Find`($G_{\alpha,\beta}$);

13 **while** $V \neq \emptyset$ **do**

14 $v \leftarrow V.front$; $V \leftarrow V \setminus \{v\}$; $G' \leftarrow h \setminus \{v\}$; $G_{\alpha,\beta} \leftarrow$ maximal $(\alpha, \beta)$-core of $G'$; $V \leftarrow V \setminus (G' \setminus G_{\alpha,\beta})$; `Find`($G_{\alpha,\beta}$);

## 3.2 A Slim Tree Structure

To address Algorithm 1's inefficiency, we propose a slim tree structure to reduce unnecessary vertex exploration.

Lemma 3.5. *For a graph G containing an $(\alpha, \beta)$-community H, if removing vertex u violates $(\alpha, \beta)$-core constraints, then there must exist a vertex $v \in N_G(u)$ not in H.*

**Proof:** Assuming all neighbors remain in $H$ after removal of $u$, $G'$ would still satisfy $(\alpha, \beta)$-core constraints, contradicting the assumption.

**Algorithm.** According to Lemma 3.5, we know that there are many redundant vertices that need to be addressed. Algorithm 2 shows the details of slim tree structure. In Algorithm 2, when $U$ is not empty, we select the first vertex of $U$ and remove it from $U$ (lines 11-12). For the current graph $h$, we delete $u$ and find its maximal $(\alpha, \beta)$-core $G_{\alpha,\beta}$ (lines 13-14) . This operation can effectively delete redundant vertices which not satisfy $(\alpha, \beta)$-core. In this way, Algorithm 2 can turn a relatively fat search tree into a slim tree. After update $U$, we We continue to call recursion (lines 15-16). Next we will give an example to describe the slim tree structure in detail.

*Example 3.6.* As Figure 3 shows, We demonstrate the process of Algorithm 2 based on Figure 2. In $S_2$, after deleting $v_6$, we need to continue deleting redundant vertices. Then we obtain $S_4$, where $u_7$, $v_7$, and $v_8$ are also deleted. Thus, in the next level of the search, the number of vertices searched decreases from 3 in to 0 , making the third level of the search tree slimmer. Similarly, in $S_3$, deleting $u_1$ simultaneously deletes $v_1$, resulting in $S_8$, which reduces the number of vertices to be searched from 5 to 4. Ultimately, we achieve the goal of pruning and significantly enhance both time and space efficiency.

Theorem 3.7. *The time complexity of Algorithm 2 is* $O((m + n + |H|(r + m)) \cdot 2^n)$.

**Proof:** Based on Algorithm 1, we performed a maximal $(\alpha, \beta)$-core search operation for each connected component, which has a time complexity of $O(m)$. Therefore, the total complexity is $O((m + n + |H|(r + m)) \cdot 2^n)$.

### 3.3 Upper Bound Algorithm

Algorithm 2 only reduces the width of the search, and the efficiency of the algorithm still cannot reach the desired effect. Therefore, we propose an upper bound-based algorithm to reduce the depth of the search as Algorithm 4 shows. The idea is that we estimate the upper bound of the weight of the current search branch. If the upper bound is smaller than the weight of the $r$-th largest influence value community found so far, we terminate the search branch. Before we introduce Algorithm 4, we will first introduce three different upper bounds.

For a connected $(\alpha, \beta)$-core $G = (U, V, E)$, where $U = \{u_1, u_2, \ldots, u_{n_u}\}$ and $V = \{v_1, v_2, \ldots, v_{n_v}\}$, for ease of expression, we denote $|U|$ as $n_u$, $|V|$ as $n_v$, $|U \cup V|$ as $n$, the maximum weight among the vertices in $U$ as $w(u)_{max}$ and the maximum weight among the vertices in $V$ as as $w(v)_{max}$.

Lemma 3.8. *Given a connected* $(\alpha, \beta)$*-core* $G = (U, V, E)$*, the first upper bound for G is defined as follows.*

$$ub_1(G) = w(u)_{max} + w(v)_{max} \quad (1)$$

**Proof:** For a connected $(\alpha, \beta)$-core $G = (U, V, E)$, we can get,

$$\begin{aligned} f(G) &= \frac{\sum_{u \in U(G)} w(u)}{n_u} + \frac{\sum_{v \in U(G)} w(v)}{n_v} \\ &\leq \frac{w(u)_{max} \cdot n_u}{n_u} + \frac{w(v)_{max} \cdot n_v}{n_v} \\ &= w(u)_{max} + w(v)_{max} \end{aligned}$$

For the subgraph $G'$ of $G$, it can also be easily proved that its upper bound is the same as that of $G$. Therefore, when the upper bound of $G$ is detected to be less than smaller than the weight of the $r$-th largest influence value community, there is no need to continue searching its subgraphs. The computational cost of $ub_1$ is cheap. It would take $O(n)$. Similarly, the next two upper bounds we will introduce follow the same logic.

---

**Algorithm 3:** Compute the upper bound

**Data:** A set $W(U) = \{w(u_1), w(u_2), \ldots, w(u_n)\}$
**Result:** Compute the upper bound
1 Shown in the [3]

---

Lemma 3.9. *Given a connected* $(\alpha, \beta)$*-core* $G = (U, V, E)$*, the second upper bound for G is defined as follows.*

$$ub_2(G) = \frac{\sum_{u \in U(G)} w(u)}{\beta} + \frac{\sum_{v \in U(G)} w(v)}{\alpha} \quad (2)$$

**Proof:** For a connected $(\alpha, \beta)$-core $G = (U, V, E)$, we can easily conclude that $n_u \geq \beta$ and $n_v \geq \alpha$. So we can get,

$$\begin{aligned} f(G) &= \frac{\sum_{u \in U(G)} w(u)}{n_u} + \frac{\sum_{v \in U(G)} w(v)}{n_v} \\ &\leq \frac{\sum_{u \in U(G)} w(u)}{\beta} + \frac{\sum_{v \in U(G)} w(v)}{\alpha} \end{aligned}$$

Similarly, For a subgraph $G'$ of $G$, if $G'$ is also a connected $(\alpha, \beta)$-core, then we can get,

$$\begin{aligned} f(G') &= \frac{\sum_{u \in U(G')} w(u)}{n'_u} + \frac{\sum_{v \in U(G')} w(v)}{n'_v} \\ &\leq \frac{\sum_{u \in U(G)} w(u)}{n'_u} + \frac{\sum_{v \in U(G)} w(v)}{n'_v} \\ &\leq \frac{\sum_{u \in U(G)} w(u)}{\beta} + \frac{\sum_{v \in U(G)} w(v)}{\alpha} \end{aligned}$$

The computational cost of $ub_2$ is also cheap. It would take $O(n)$.

The second upper bound would only be tight when $G$ contains an optimum result with the size of $U$ close to $\beta$ and the size of $V$ close to $\alpha$. However, it has limited pruning effectiveness when $G$ contains large-size results. Next we study tight bounds for arbitrary $G$.

Lemma 3.10. *Given a connected* $(\alpha, \beta)$*-core* $G = (U, V, E)$*, the third upper bound for G is defined as follows.*

$$ub_3(G) = max\ \{f(S) | S \subseteq G\} \quad (3)$$

**Proof:** For a connected $(\alpha, \beta)$-core $G = (U, V, E)$, let $S^*$ be the $(\alpha, \beta)$-influential community in $G$ and $S$ be $\max\{f(S)|S \subseteq G\}$. According to the definition, $S^*$ must satisfy $(\alpha, \beta)$-core, however, $S'$ relaxes this constraint, so $f(S^*) \leq f(S')$ must be hold.

The computational cost of $ub_3$ is expensive. If using exhaustive method, it would take $O(2^{n+m})$. However, there is a simple and effective approximate algorithm [3] that can achieve (1/2)-approximation with complexity $O(n)$. As such we can use the approximation algorithm to get an at least 1/2 $ub_3(G)$ value first and then multiple it by 2 to derive a slightly loose bound. The specific calculation of $ub_3$ is shown in Algorithm 3.

In Algorithm 3, all vertices of bipartite graph $G$ are denoted by $U$, with weights stored in $W(U)$. The algorithm initializes two sets (line 1), then computes the impact of adding or removing each $w(u_i)$ (lines 3-4). Based on $a'_i$ and $b'_i$, it makes greedy choices (lines

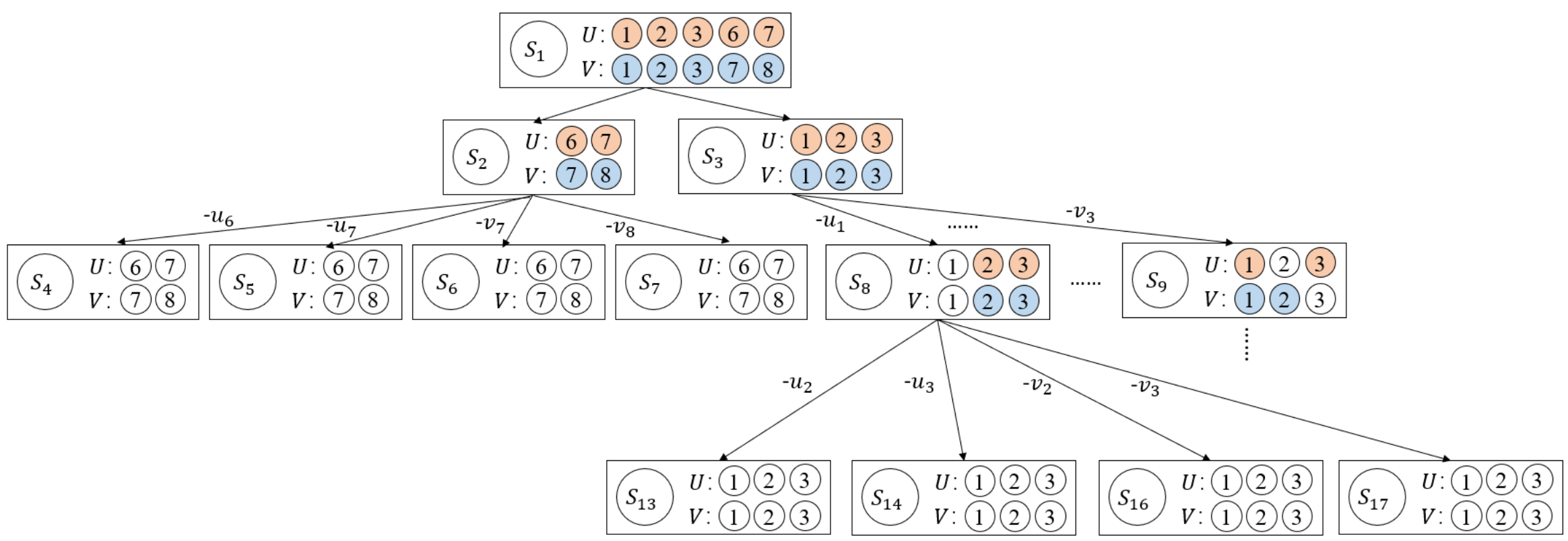

**Figure 3: An example of a slim tree structure**

**Algorithm 4:** Upper Bound Algorithm

**Input:** $G = (U, V, E)$, $\alpha$, $\beta$, $r$
**Output:** Top-$r$ ($\alpha$,$\beta$)-influential communities

1 **Function** `Main()`:
2 $S \leftarrow \emptyset$; `Find`$(G)$; **return** $S$;

3 **Procedure** `Find`$(G)$:
4 $G \leftarrow$ maximal $(\alpha, \beta)$-core of $G$; $H \leftarrow$ connected components of $G$; $h_{min} \leftarrow r$-th largest community in $S$; **foreach** $h \in H$ **do**
5 **if** $f(h) > f(h_{min})$ **then**
6 $flag \leftarrow true$; **foreach** $h' \in S$ **do**
7 **if** $f(h) = f(h')$ ***and*** $h \subseteq h'$ **then**
8 $flag \leftarrow false$; **break;**
9 **if** $flag$ **then**
10 $S \leftarrow (S \setminus h_{min}) \cup h$;
11 **while** $U \neq \emptyset$ **do**
12 $u \leftarrow U.front$; $U \leftarrow U \setminus \{u\}$; $G' \leftarrow h \setminus \{u\}$; $G_{\alpha,\beta} \leftarrow$ maximal $(\alpha, \beta)$-core of $G'$; $U \leftarrow U \setminus (G' \setminus G_{\alpha,\beta})$; $L \leftarrow$ three upper bounds of $G_{\alpha,\beta}$; $ub \leftarrow \min(L)$; **if** $ub > h_{min}$ **then**
13 `Find`$(G_{\alpha,\beta})$;
14 **while** $V \neq \emptyset$ **do**
15 $v \leftarrow V.front$; $V \leftarrow V \setminus \{v\}$; $G' \leftarrow h \setminus \{v\}$; $G_{\alpha,\beta} \leftarrow$ maximal $(\alpha, \beta)$-core of $G'$; $V \leftarrow V \setminus (G' \setminus G_{\alpha,\beta})$; $L \leftarrow$ three upper bounds of $G_{\alpha,\beta}$; $ub \leftarrow \min(L)$; **if** $ub > h_{min}$ **then**
16 `Find`$(G_{\alpha,\beta})$;

6-12) and finally returns avg$(X[n])$. A formal (1/2)-approximation proof is provided in [3] and omitted here.

**Algorithm.** The difference between Algorithm 4 and Algorithm 2 lies in the reduction of the search depth. For the current graph being searched, three upper bounds are calculated, and the tightest upper bound is compared with $h_{min}$. If it is greater than $h_{min}$, the search continues; otherwise, the current branch is terminated.Therefore, The total number of recursions in Algorithm 4 is significantly reduced compared to Algorithm 2, greatly reducing both space and time requirements.

*Example 3.11.* As Figure 4 shows, We demonstrate the process of Algorithm 4. We find that the community with the greatest influence is $S_2$. Next, we process $S_3$ and delete $u_1$, resulting in $S_8$. We discover that the upper bound of the influence value of the current graph is smaller than that of $S_2$, so we do not need to perform the next level of search. This approach reduces the depth of the search.

THEOREM 3.12. *The time complexity of Algorithm 4 is* $O((m + n + |H|(r + m)) \cdot 2^n)$.

**Proof:** Based on the process of Algorithm 4, we can easily conclude that the time complexity of this algorithm is the same as that of Algorithm 2.

# 4 APPROXIMATE ALGORITHMS

Due to the time-consuming nature of exact algorithms, we propose a heuristic algorithm in this section. The algorithm is based on a greedy strategy and aims to find a sufficiently good solution within a reasonable time frame, thus ensuring efficiency and practicality.

## 4.1 New Framework

The heuristic algorithm employs a greedy strategy, selecting the highest-weight vertex iteratively to build an $(\alpha, \beta)$-influential community.

**Algorithm.** Algorithm 5 begins similarly to previous algorithms (lines 6-8). For each connected component, it initializes an empty queue $Q$ and an empty graph $G'$ (line 10). The algorithm selects and enqueues the maximum-weight vertex from $U$ (lines 11-12). Vertices dequeued from $Q$ are added to $G'$ and marked as visited (lines 13-16). For each dequeued vertex $v \in U$, its neighbors are sorted by weight and the top $\gamma$ are selected using the *Check* function (lines 17-21). Unvisited neighbors are then enqueued (lines 22-24). The process is analogous for $v \in V$. The procedure continues until

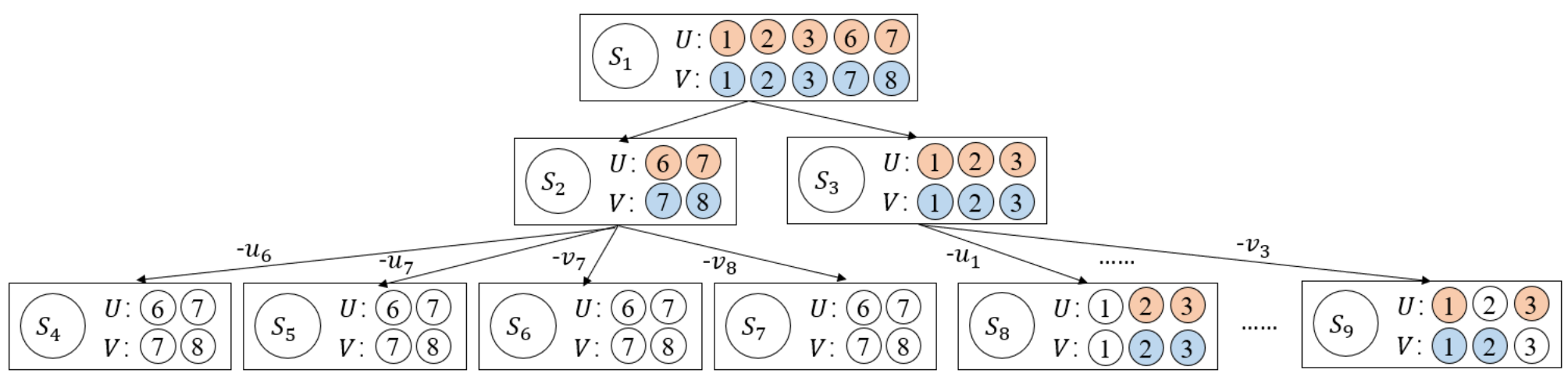

**Figure 4: An example of upper bound algorithm**

the queue is empty, after which $G'$ is added to the solution set $S$ (lines 27-28).

*Example 4.1.* Figure 5 demonstrates Algorithm 5, identifying a top-1 (2,2)-influential community. Starting from the highest-weight vertex $u_6$, vertices are iteratively added based on their weights and connections until forming the final community $S_7$.

**Algorithm 5:** New Framework

**Input:** $G = (U, V, E), \alpha, \beta, r$

**Output:** Top-$r$ $(\alpha,\beta)$-influential communities

1 **Function** `Main()`:

2 $S \leftarrow \emptyset$; `Find`$(G)$; **return** $S$;

3 **Procedure** `Find`$(G)$:

4 $G \leftarrow$ maximal $(\alpha, \beta)$-core of $G$; $H \leftarrow$ connected components of $G$; $h_{min} \leftarrow r$-th largest influence community in $S$; **foreach** $h \in H$ **do**

5 $Q \leftarrow \emptyset, G' \leftarrow \emptyset$; $u \leftarrow$ max-weight vertex in $U(h)$; $Q.push(u)$; **while** $Q \neq \emptyset$ **do**

6 $v \leftarrow Q.pop()$; $G'.add(v)$; $vis[v] \leftarrow true$; **if** $v \in U(h)$ **then**

7 $N \leftarrow$ neighbors of $v$, sorted by descending weight; $\gamma \leftarrow$ `Check`$(N)$; $N \leftarrow$ top $\gamma$ vertices; **foreach** $u' \in N$ **do**

8 **if** $!vis[u']$ **then**

9 $Q.push(u')$

10 **if** $v \in V(h)$ **then**

11 lines 9-14 with $\alpha$ and $\beta$ swapped;

12 **if** $G'$ *is* $(\alpha, \beta)$*-core and* $f(G') > f(h_{min})$ **then**

13 $S \leftarrow (S \setminus h_{min}) \cup G'$;

14 **Procedure** `Check`$(N)$:

15 $num \leftarrow 0$; $u_0 \leftarrow$ first vertex in $N$; **foreach** $u \in N$ **do**

16 **if** $w(u) = w(u_0)$ **then**

17 $num++$;

18 **else**

19 break;

20 **return** *max*$(num, \alpha)$;

Theorem 4.2. *Algorithm 5 correctly identifies and finds the $(\alpha, \beta)$-influential communities.*

**Proof:** We will discuss the correctness of Algorithm 5. We ensure the connectivity of the community by adding neighboring vertices each time, and after each addition, we determine whether it is an $(\alpha, \beta)$-core to ensure the cohesiveness of the community. Finally, we will specifically explore how to satisfy the maximality constraint. Assuming the current graph $G$, since the weight of each vertex we add is non-increasing, the new graph $G'$ obtained will definitely have $f(G') \leq f(G)$. $f(G') = f(G)$ if and only if $f_U(G') = f_U(G)$,$f_V(G') = f_V(G)$. Therefore, when $f(G') = f(G)$, the weights of the upper layer vertices in G' are all equal, and the weights of the lower layer vertices are all equal. At this time, the maximality constraint is not satisfied. Therefore, we ensure the maximality constraint through the *Check* function. If the number of the largest values is greater than $\alpha$, then all the largest values are added to the queue $Q$. Otherwise, only $\alpha$ of them are added. This ensures the maximum constraint.

Theorem 4.3. *The time complexity of Algorithm 5 is $O((n+m) + m \log n)$.*

**Proof:** In Algorithm 5, assume graph $G$ has $n$ vertices and $m$ edges. First, calculating the maximal $(\alpha, \beta)$-core has a time complexity of $O(m)$, followed by the need to find connected components, which requires $O(n + m)$ time. Assuming that sorting is needed for the neighbors of each vertex, the time complexity reaches $O(m \log n)$. Therefore, the overall time complexity of Algorithm 5 is $O((n + m) + m \log n)$.

## 4.2 Pruning Algorithm

**Algorithm.** Based on Algorithm 5, we propose Pruning Algorithm 6. The difference in Algorithm 6 is in line 16, where we make an influence value judgment. Suppose the current graph is $G$, the influence value of the new graph $G'$ created after adding a new vertex in $G$ is guaranteed to be less than or equal to $G$, that is, $f(G') \leq f(G)$. If $f(G) < f(h_{min})$, then the influence value of any subsequent graphs explored will definitely be less than $f(h_{min})$. Therefore, we will not search the subsequent vertices anymore, thereby achieving effective pruning.

*Example 4.4.* Based on Algorithm 5, we show the process of Algorithm 6 in Figure 5. After obtaining $S_9$, we find that the current graph's influence value is 4, which is smaller than the influence value of $S_7$ which has the largest influence value. Since the weight of the added vertices is non-increasing, the influence of the subsequent graphs will certainly be less than or equal to 4. Therefore, we do not need to continue searching the subsequent graphs.

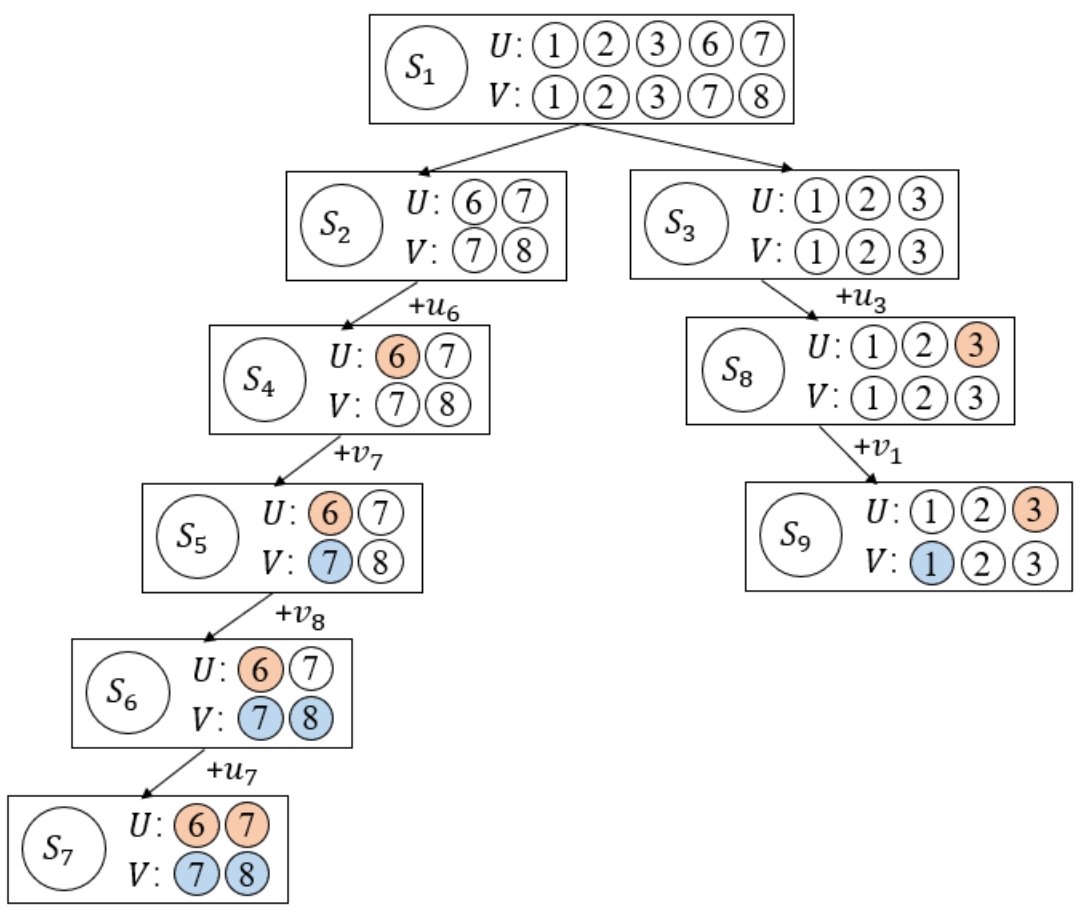

**Figure 5: An example of pruning algorithm**

THEOREM 4.5. *The time complexity of Algorithm 6 is* $O((n + m) + m \log n)$.

**Proof:** Based on Algorithm 5, Algorithm 6 adds an $O(1)$ operation (lines 16-17), so the time complexity remains unchanged, which is $O((n + m) + m \log n)$.

## 5 EXPERIMENTS

This section presents our experimental results. All algorithms are implemented in C++. All experiments are performed under a Linux operating system on a machine with an Intel Xeon Platinum 8373C 2.6GHz CPU and 188G memory. In this set of experiments, we set the maximum running time for each test as 1 hour. If a test does not stop within the time limit, we denote its processing time as INF.

**Datasets**. We use 10 real-world bipartite graphs from KONECT (http://konect.cc/). Table 1 summarizes the datasets: $|U|$, $|V|$, and $|E|$ denote the numbers of upper-layer vertices, lower-layer vertices, and edges, respectively. $d_U^{max}$ and $d_V^{max}$ indicate the maximum degrees in $U$ and $V$. Since original datasets lack vertex weights, we assign weights via uniform distribution.

**Exact Algorithms** :

- Baseline: Basic top-$r$ $(\alpha, \beta)$-influential community search (Algorithm 1).
- SlimTree: Slim tree structure to reduce redundancy (Algorithm 2).
- UpperBound: Search pruning using upper bounds (Algorithm 4).

**Approximate Algorithms** :

- NewFra: Greedy-based approximation method (Algorithm 5).
- Pruning: Approximation via pruning strategy (Algorithm 6).

### 5.1 Experiments of Exact Algorithms

**Exp-I** : **Varying** $\alpha(\beta)$. Figure 6 shows that as $\alpha$ increases, running time decreases due to reduced search space. UpperBound outperforms Baseline and SlimTree significantly, especially for larger graphs (e.g., at $\alpha = 7$, over 100x faster). Results for varying $\beta$ (Figure 7) show similar trends.

**Algorithm 6:** Pruning Algorithm

**Input:** $G = (U, V, E)$, $\alpha$, $\beta$, $r$
**Output:** Top-$r$ $(\alpha,\beta)$-influential communities

1 **Function** Main():
2 $\quad S \leftarrow \emptyset$; Find($G$); **return** $S$;
3 **Procedure** Find($G$):
4 $\quad G \leftarrow$ maximal $(\alpha, \beta)$-core of $G$; $H \leftarrow$ connected components of $G$; $h_{min} \leftarrow r$-th largest influence community in $S$; **foreach** $h \in H$ **do**
5 $\quad\quad Q \leftarrow \emptyset$, $G' \leftarrow \emptyset$; $u \leftarrow$ max-weight vertex in $U(h)$; $Q.push(u)$; **while** $Q \neq \emptyset$ **do**
6 $\quad\quad\quad v \leftarrow Q.pop()$; $G'.add(v)$; **if** $f(G') < h_{min}$ *and* $V(G') \neq \emptyset$ **then**
7 $\quad\quad\quad\quad$ break;
8 $\quad\quad\quad vis[v] \leftarrow true$; **if** $v \in U(h)$ **then**
9 $\quad\quad\quad\quad N \leftarrow$ neighbors of $v$, sorted by descending weight; $\gamma \leftarrow$ Check($N$); $N \leftarrow$ top $\gamma$ vertices; **foreach** $u' \in N$ **do**
10 $\quad\quad\quad\quad\quad$ **if** $!vis[u']$ **then**
11 $\quad\quad\quad\quad\quad\quad Q.push(u')$
12 $\quad\quad\quad$ **if** $v \in V(h)$ **then**
13 $\quad\quad\quad\quad$ lines 9-14 with $\alpha$ and $\beta$ swapped;
14 $\quad\quad$ **if** $G'$ *is* $(\alpha, \beta)$*-core and* $f(G') > f(h_{min})$ **then**
15 $\quad\quad\quad S \leftarrow (S \setminus h_{min}) \cup G'$;

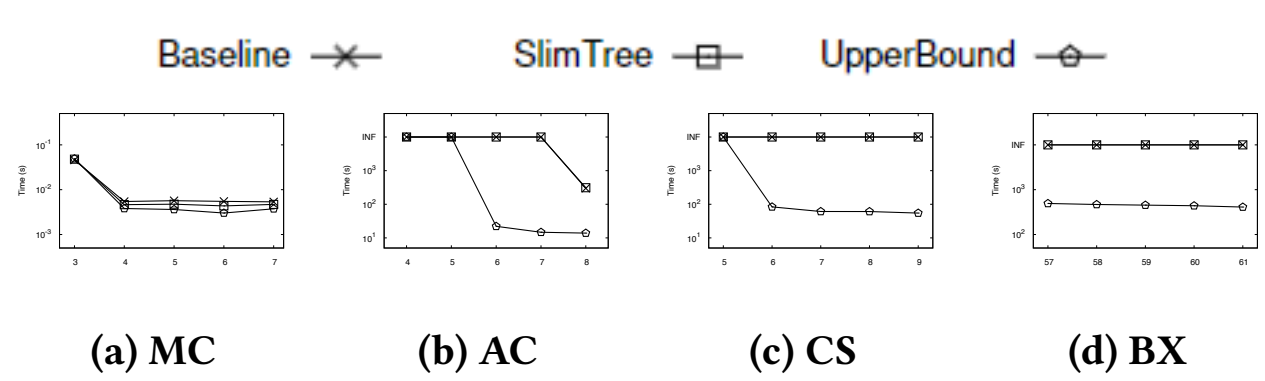

(a) MC (b) AC (c) CS (d) BX

**Figure 6: Running time of Exact algorithms (Vary $\alpha$)**

**Table 1: Summary of Datasets**

| Dataset | $\lvert U\rvert$ | $\lvert V\rvert$ | $\lvert E\rvert$ | $\alpha_{max}$ | $\beta_{max}$ |
| --- | --- | --- | --- | --- | --- |
| MC | 0.8K | 0.6K | 1.5K | 25 | 18 |
| AC | 17K | 22K | 59K | 116 | 18 |
| MA | 6.5K | 19K | 100K | 1,625 | 111 |
| GH | 56K | 120K | 440K | 884 | 3,675 |
| CS | 105K | 181K | 512K | 286 | 385 |
| BX | 105K | 340K | 1.15M | 13,601 | 2,502 |
| DBT | 64K | 88K | 3.23M | 6,507 | 12,400 |
| PA | 1.95M | 5.62M | 12.28M | 1,386 | 1,386 |
| LG | 3.21M | 7.49M | 112.31M | 300 | 1,053,676 |
| WT | 27.67M | 12.76M | 140.61M | 1,100,065 | 11,571,952 |

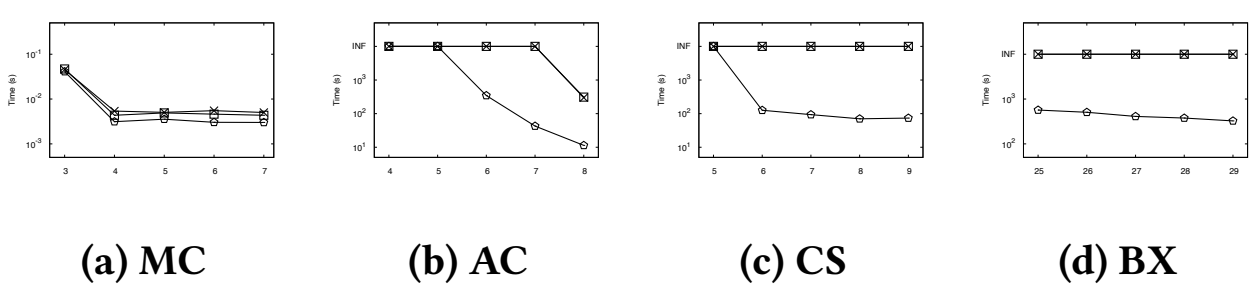

(a) MC (b) AC (c) CS (d) BX

**Figure 7: Running time of Exact algorithms (Vary $\beta$)**

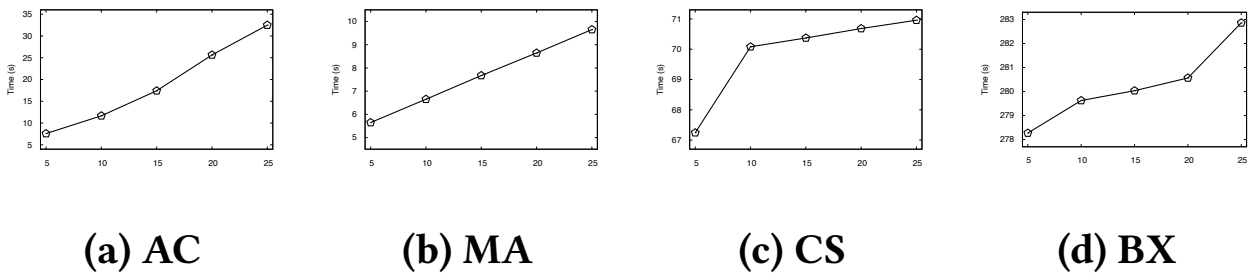

(a) AC (b) MA (c) CS (d) BX

**Figure 8: Running time of UpperBound (Vary $r$)**

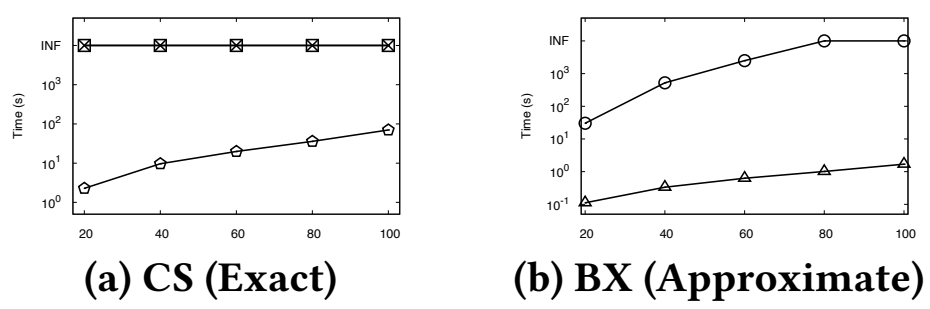

(a) CS (Exact) (b) BX (Approximate)

**Figure 9: Scalability of Algorithms**

**Exp-II** : **Varying r**. Only UpperBound is evaluated due to timeouts from other algorithms. As $r$ increases, runtime increases moderately (Figure 8) due to more required outputs and iterations.

**Exp-III** : **Scalability of Exact Algorithms**. Figure 9a shows that UpperBound scales well from 2s to 70s with increasing data, while Baseline and SlimTree fail on larger samples. This demonstrates UpperBound's superior scalability.

## 5.2 Experiments of Approximate Algorithms

**Exp-IV** : **Varying** $\alpha(\beta)$ : Figure 10 shows Pruning consistently outperforms NewFra (1–3 orders of magnitude). The performance gap widens with larger datasets due to NewFra's full-graph traversal vs. Pruning's effective pruning. At $\alpha = 96$, NewFra fails while Pruning completes in 405s. As $\alpha$ increases, both algorithms run faster due to reduced graph size. Similar trends appear for $\beta$ (Figure 11). Notably, in BX dataset, Pruning takes 0.4s while UpperBound takes 492s, confirming Pruning's efficiency.

**Exp-V** : **Varying r**. Figures 12 and 13 show runtime increases slightly with $r$. The effect is minimal, often under 0.01s, indicating $r$ has limited impact when small.

**Exp-VI** : **Scalability of Approximate Algorithms**. Figure 9b confirms both NewFra and Pruning scale near-linearly. Pruning remains about 2 orders of magnitude faster, supporting previous findings.

## 6 CONCLUSION

In this paper, we introduce the ($\alpha$,$\beta$)-influential community model. We define the influence of a community as the sum of the average weights of the upper-layer vertices and the lower-layer vertices, thereby comprehensively reflecting the community's influence. To find ($\alpha$,$\beta$)-influential communities, we propose an exact algorithm and optimized it. Due to the time-consuming nature of exact algorithms, we propose approximate algorithms which only take $O((n + m) + m \log n)$ time. The efficiency and the effectiveness of our proposed techniques are verified through extensive experiments.

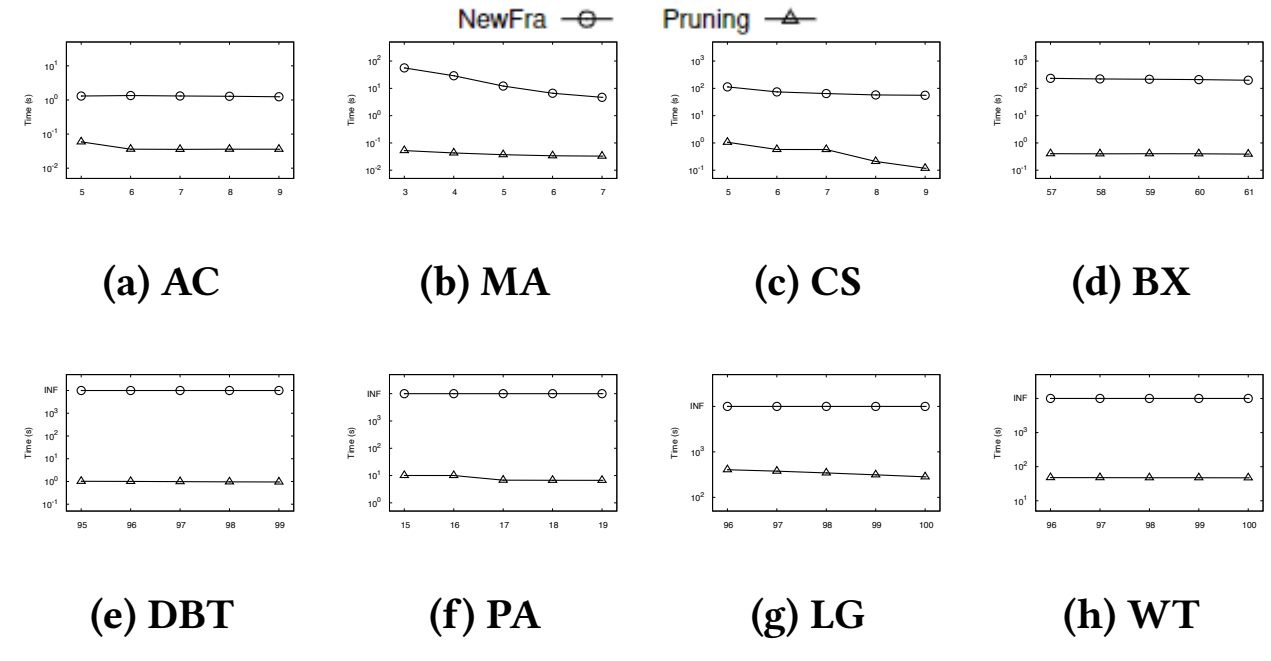

(a) AC (b) MA (c) CS (d) BX

(e) DBT (f) PA (g) LG (h) WT

**Figure 10: Running time of Approximate algorithms (Vary $\alpha$)**

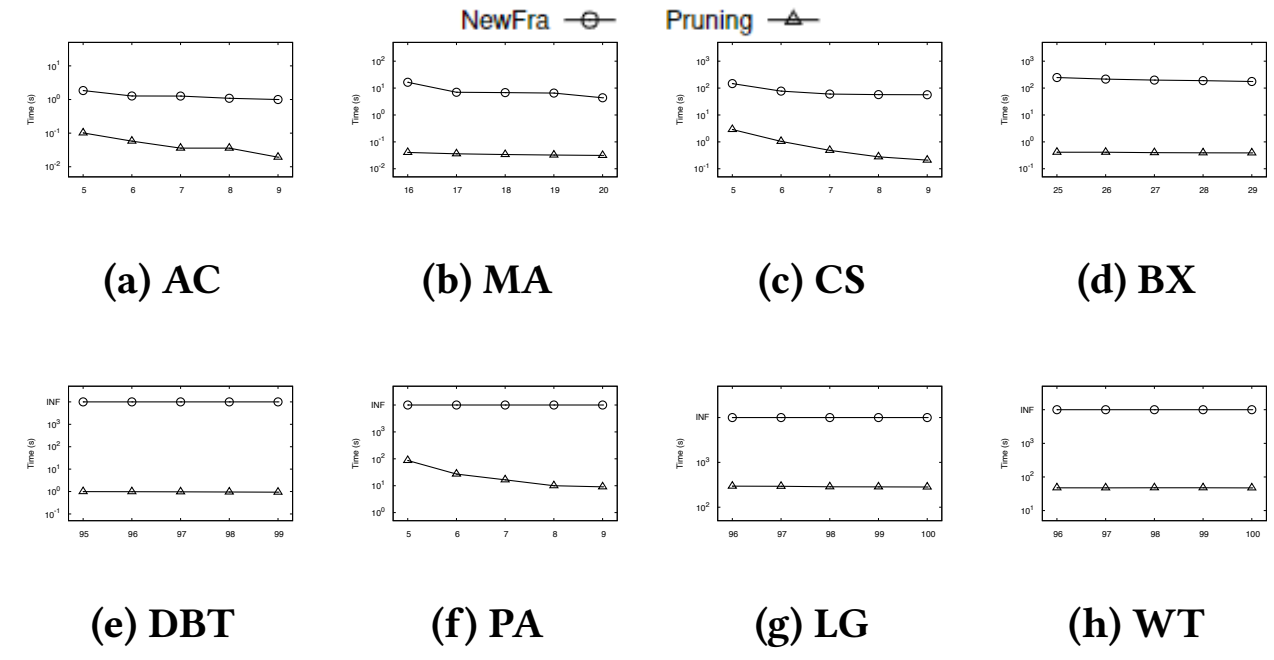

(a) AC (b) MA (c) CS (d) BX

(e) DBT (f) PA (g) LG (h) WT

**Figure 11: Running time of Approximate algorithms (Vary $\beta$)**

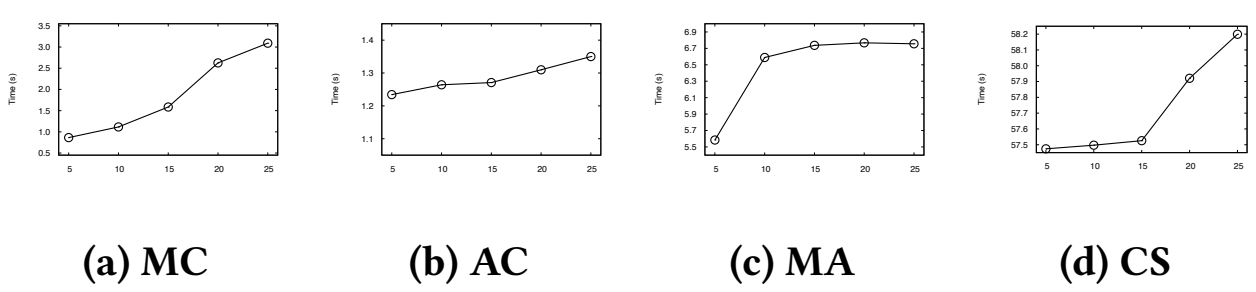

(a) MC (b) AC (c) MA (d) CS

**Figure 12: Running time of Newfra (Vary $r$)**

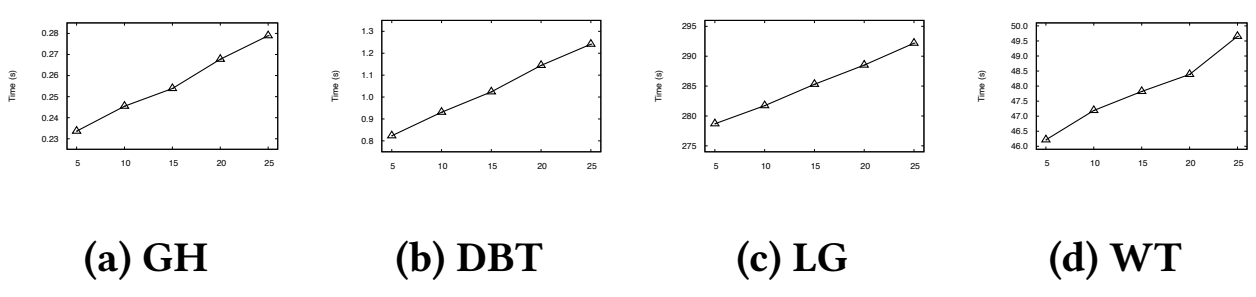

(a) GH (b) DBT (c) LG (d) WT

**Figure 13: Running time of Pruning (Vary $r$)**